\documentstyle[12pt]{article}
\def\a0{\bar\alpha_0}
\def\ae{\alpha_{\mbox{\tiny S}}^{\mbox{\scriptsize eff}}}
\def\ap{\bar\alpha_p}
\def\as{\alpha_{\mbox{\tiny S}}}
\def\b0{\beta_0}
\def\cd{\chi^2/\mbox{d.o.f.}}
\def\ee{e^+e^-}
\def\ex{{\mbox{\scriptsize exp}}}
\def\kt{k_\bot}

\def\mI{\mu_{\mbox{\tiny I}}}
\def\mR{\mu_{\mbox{\tiny R}}}
\def\MSbar{\overline{\mbox{MS}}}
\def\pt{{\mbox{\scriptsize pert}}}
\def\pw{{\mbox{\scriptsize pow}}}
\def\sL{\sigma_{\mbox{\tiny L}}}
\def\st{\sigma_{\mbox{\scriptsize tot}}}
\def\th{{\mbox{\scriptsize th}}}
\def\frac#1#2{ {{#1} \over {#2} }}

\def\VEV#1{\left\langle #1\right\rangle}
\def\beq{\begin{equation}}
\def\beeq{\begin{eqnarray}}
\def\eeq{\end{equation}}
\def\eeeq{\end{eqnarray}}
\def\np#1#2#3{Nucl.\ Phys.\ B#1 (19#3) #2}
\def\pl#1#2#3{Phys.\ Lett.\ #1B (19#3) #2}
\def\pr#1#2#3{Phys.\ Rev.\ D#1 (19#3) #2}
\def\prl#1#2#3{Phys.\ Rev.\ Lett.\ #1 (19#3) #2}
\def\zp#1#2#3{Z.\ Phys.\ C#1 (19#3) #2}
\begin{document}
\begin{titlepage}
\renewcommand{\thefootnote}{\fnsymbol{footnote}}
\begin{flushright}
     Cavendish--HEP--95/2 \\
     LU TP 95--8 \\
     hep-ph/9504219
\end{flushright}
\vspace*{\fill}
\begin{center}
{\Large \bf
Calculation of Power Corrections\\[1ex]
to Hadronic Event Shapes\footnote{Research supported in part by
the U.K. Particle Physics and Astronomy Research Council and by
the EC Programme ``Human Capital and Mobility", Network ``Physics at
High Energy Colliders", contract CHRX-CT93-0357 (DG 12 COMA).}}
\end{center}
\par \vskip 2mm
\begin{center}
        {\bf Yu.L.\ Dokshitzer\footnote{Permanent address:
St Petersburg Nuclear Physics Institute, Gatchina, St Petersburg 188350,
Russian Federation}} \\
         Department of Theoretical Physics, University of Lund, \\
         S\"olvegatan 14 A, S-223 62 Lund, Sweden \\[1ex]
             and \\[1ex]
        {\bf B.R.\ Webber} \\
        Cavendish Laboratory, University of Cambridge,\\
        Madingley Road, Cambridge CB3 0HE, U.K.
\end{center}
\par \vskip 2mm
\begin{center} {\large \bf Abstract} \end{center}
\begin{quote}
We compute power corrections to hadronic event shapes in
$\ee$ annihilation, assuming an infrared regular behaviour
of the effective coupling $\as$. With the integral of $\as$
over the infrared region as the only non-perturbative parameter,
also measured in heavy quark physics, we can account for
the empirical features of $1/Q$ corrections to the mean
values of various event shapes.
\end{quote}
\vspace*{\fill}
\begin{flushleft}
     Cavendish--HEP--95/2\\
     LU TP 95--8 \\
     March 1995
\end{flushleft}
\end{titlepage}
\renewcommand{\thefootnote}{\fnsymbol{footnote}}

\noindent{\large\bf 1. Introduction}

\par \vskip 1ex\noindent
Infrared-safe shape measures for hadronic final states in
$\ee$ annihilation would appear in principle to be an
ideal testing ground for perturbative QCD. Such quantities
are asymptotically insensitive to long-distance non-perturbative
physics and can thus be computed order-by-order in perturbation
theory. The large momentum scale $Q\sim M_Z$ available at
existing $\ee$ machines implies a small value of the
running coupling $\as(Q)$ and so the perturbation series
should be relatively well-behaved. Non-perturbative
effects should be suppressed by inverse powers of $Q$.
Thus hadronic event shapes should be capable of providing
a high-precision measurement of $\as$.

Unfortunately the hoped-for precision has not yet been
achieved, partly because ${\cal O}(\as^3)$ calculations
of event shapes are still lacking, but also because
non-perturbative effects turn out to be significant
even at $Q\sim M_Z$. This is because they are in fact
suppressed by only a single inverse power of $Q$ in
many cases. Bearing in mind that $\as(M_Z)\sim 0.12$
and the non-perturbative scale is ${\cal O}$(1 GeV),
we see that the power correction may easily be comparable
with the ${\cal O}(\as^2)$ next-to-leading term at
present energies. Consequently it becomes essential
to achieve some understanding of power corrections
before embarking on any ${\cal O}(\as^3)$ calculations
of event shapes.

In the present paper we adopt the approach, advocated in
Refs.~[\ref{DKTheavy},\ref{hadro}], of trying to deduce as
much as possible about power corrections from perturbation
theory. In particular we explore the consequences of assuming
that $\as$, defined in some appropriate way, does not
grow indefinitely at low scales but instead has an
infrared-regular effective form. Then various moments
of $\as$, integrated over the infrared region, play the
r\^ole of non-perturbative parameters which determine
the form and magnitude of power corrections. Since
$\as$ is supposed to be universal, we obtain relations
between the power corrections to various observables.

Our approach is related to that of Korchemsky and
Sterman [\ref{KS}], and also to several other recent
papers that discuss power corrections and the ambiguities
of perturbation theory in terms of infrared
renormalons [\ref{mueller}], in the context of
the Drell-Yan process [\ref{CS}], event shapes [\ref{MW}],
deep inelastic scattering [\ref{rendis}], heavy quark
effective theory [\ref{renhvy}] or quark
confinement [\ref{renconf}]. From our viewpoint,
infrared renormalons arise from the divergence of
the perturbative expression for $\as$ at low scales,
and the ambiguities associated with different ways of
avoiding the renormalon poles in the Borel transform
plane are resolved by specifying the infrared
behaviour of $\as$.  This approach implies
relationships between the contributions of
a given renormalon to different processes.

The quantitative results we obtain look very good
in the case of the mean value of the thrust parameter
[\ref{thrust}]. The required value of the relevant
moment of $\as$ is consistent with that obtained
from a similar approach to heavy quark
fragmentation [\ref{DKTheavy}].  For the
other shape variables discussed here, the
mean value of the $C$-parameter [\ref{cpar}]
and the longitudinal cross section [\ref{sigl}],
a comparison with LEP data is encouraging, but
detailed tests must await the re-analysis of
lower-energy data to establish the energy
dependence of the leading power correction.

\newpage
\noindent{\large\bf 2. Calculations}

\par \vskip 1ex\noindent
Consider a quantity of the form
\beq\label{Fdef}
F = \int_0^Q dk\,f(k)
\eeq
where $f(k)$ behaves like $\as(k)\,k^p$ at $k\ll Q$, say
\beq
f(k) \sim a_F\,\as(k)\,k^p/Q^{p+1}\;\;\;\;\;\;\;(k\ll Q)
\eeq
where we have included the appropriate $Q$ dependence assuming
$F$ is dimensionless.  Suppose that $F$ has the perturbative expansion
\beq
F^\pt = F_1\,\as + F_2\,\as^2 + \cdots\;.
\eeq
More precisely, if the coefficients $F_n$ are computed in the
$\MSbar$ renormalization scheme at scale $Q$, then in terms of
the $\MSbar$ coupling at scale $\mR$ we have
\beq\label{Fpt}
F^\pt = F_1\,\as(\mR) +\left(F_2 + \frac{\b0}{2\pi}\ln\frac{\mR}{Q}F_1\right)
\as^2(\mR) + \cdots
\eeq
where $\b0=(11C_A -2N_f)/3$, with $C_A=3$, for $N_f$ active flavours.

In part of the integration region of Eq.~(\ref{Fdef}) the
perturbative expression for $\as(k)$ is not appropriate.
We may however choose an infrared matching
scale $\mI$ such that $\Lambda\ll \mI\ll Q$ and assume that perturbation
theory is valid above that scale. We can then introduce a non-perturbative
parameter $\ap(\mI)$ to represent the portion of the integral below $\mI$:
\beq\label{apdef}
\int_0^{\mI} dk\,\as(k)\,k^p \equiv \frac{\mI^{p+1}}{p+1}\,\ap(\mI)\;.
\eeq
Before adding this contribution to $F^\pt$, we have to subtract the
perturbative value of this integral, which is, to second order,
\beq\label{ptint}
\frac{\mI^{p+1}}{p+1}\left[\as(\mR) +\frac{\b0}{2\pi}\left(
\ln\frac{\mR}{\mI}+\frac{1}{p+1}\right)\as^2(\mR) \right]\;.
\eeq

As a refinement, and for consistency with Ref.~[\ref{DKTheavy}], we
shall assume that the parameter $\ap(\mI)$ refers not to the coupling
in the $\MSbar$ scheme but rather to the scheme proposed in
Ref.~[\ref{CMW90}],
which is expected to be more physical in the region under consideration.
Thus $\as$ in Eq.~(\ref{apdef}) (only) is to be interpreted as $\ae$
where in terms of the $\MSbar$ coupling
\beq
\ae = \as + \frac{K}{2\pi}\as^2
\eeq
with
\beq
K = \left(\frac{67}{18}-\frac{\pi^2}{6}\right)C_A-\frac{5}{9}N_f\;.
\eeq
The only effect on Eq.~(\ref{ptint}) is that the term $\ln(\mR/\mI)$
becomes $\ln(\mR/\mI) + K/\b0$. The full expression for $F$ is thus
\beq
F = F^\pt + F^\pw
\eeq
where $F^\pt$ is as given in Eq.~(\ref{Fpt}) while
\beq\label{Fpow}
F^\pw = \frac{a_F}{p+1}\left(\frac{\mI}{Q}\right)^{p+1}
\left[\ap(\mI) - \as(\mR) - \frac{\b0}{2\pi}\left(
\ln\frac{\mR}{\mI}+\frac{K}{\b0} +\frac{1}{p+1}\right)\as^2(\mR)\right]\;.
\eeq
The dependence of $F^\pt$ on the renormalization scale $\mR$ is
one order higher in $\as$ than that calculated, i.e.\ third-order
in this case. Similarly, the dependence of the power correction
$F^\pw$ on both $\mR$ and the infrared matching
scale $\mI$ is third-order, provided $\mI$ is
sufficiently large for $\as(\mI)$ to have reached its perturbative
form. Of course, the value obtained for $\ap(\mI)$ depends on $\mI$,
but this is mostly compensated by the other $\mI$-dependent term.

The value of the power $p$ and the coefficient $a_F$ may be found from
the infrared cutoff dependence of the lowest-order perturbative result.
In this connection, it is crucial that the appropriate argument of $\as$
for soft and/or collinear gluon emission is the gluon transverse
momentum $\kt$ [\ref{askt}]. Thus the cutoff should be a $\kt$-cutoff.

Consider for example the mean value of the thrust $T$.
The contribution to this quantity from the region $\kt<\mI$ is
\beq
\delta\VEV{T} = -\frac{C_F}{2\pi}\int_{\kt<\mI}
dx_1\,dx_2\,\as(\kt)\frac{x_1^2+x_2^2}{(1-x_1)(1-x_2)}\,
\min\{(1-x_1),(1-x_2)\}
\eeq
where $C_F = 4/3$.
Setting $1-x_{1,2} = y_{1,2}$ and considering the region
$y_1<y_2\ll 1$, we have $\kt = \sqrt{y_1 y_2} Q$ and hence
\beeq
\delta\VEV{T} &=& -4\frac{C_F}{\pi}\int_0^{\mI/Q}dy_1
\int_{y_1 Q}^{\mI}\frac{d\kt}{\kt}\as(\kt) \nonumber \\
&=& -\frac{4C_F}{\pi Q}\int_0^{\mI} d\kt\,\as(\kt)
\;\equiv\; -4\frac{C_F}{\pi}\,\frac{\mI}{Q}\,\a0(\mI)\;.
\eeeq
Thus in this case $p=0$ and we obtain a $1/Q$ correction, with
a coefficient in Eq.~(\ref{Fpow}) of $a_F = a_T$ where
\beq\label{aT}
a_T = -4\frac{C_F}{\pi} = -1.70\;.
\eeq

As shown by the solid curve in Fig.~1, an excellent fit to the data
on $\VEV{T}$ over the range $14<Q<92$ GeV can be obtained using the
perturbative prediction [\ref{lepqcd}]
\beq\label{Tpert}
\VEV{T}^\pt = 1-0.335\,\as - 1.02\,\as^2
\eeq
with $\mR=Q$ and $\as(M_Z) = 0.117\pm 0.005$ [\ref{glasgow}], plus a
power correction of the form (\ref{Fpow}). For $\mI=2$ GeV, the
fitted value of the non-perturbative parameter $\a0$ is
\beq\label{a0}
\a0(2\;\mbox{GeV}) \equiv (\mbox{2 GeV})^{-1}
\int_0^{\mbox{\scriptsize 2 GeV}} dk\, \ae(k)
= 0.53\pm 0.04\;,
\eeq
with $\chi^2 = 4.5$ for 8 degrees of freedom.
Allowing both $\as(M_Z)$ and $\a0(2\;\mbox{GeV})$ to
be free parameters gives
\beq
 \as(M_Z) = 0.120\pm 0.004\;,\;\;\;\;\a0(2\;\mbox{GeV})= 0.52\pm 0.03\;,
\eeq
with $\cd = 3.7/7$.

\begin{figure}[htb]
\vspace{11.5cm}
\includegraphics{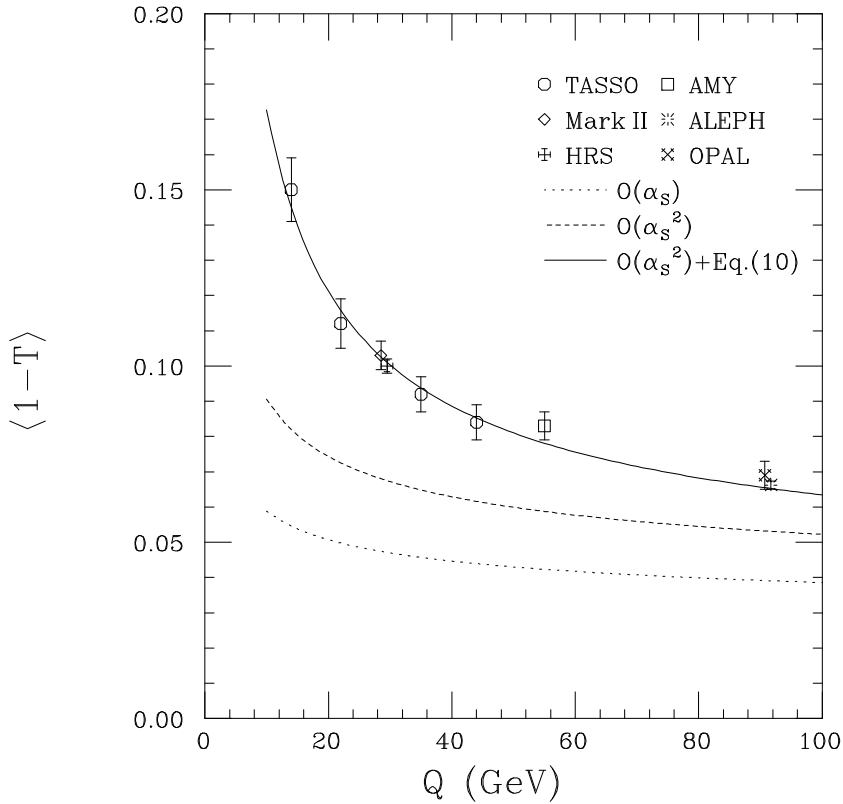}
\caption{Mean value of $1 - T$, where $T$ is the thrust.}\label{meant}
\end{figure}

We also obtain good fits for other values of the arbitrary infrared
matching parameter $\mI$. At $\mI=3$ GeV, for example, we find
\beq
 \as(M_Z) = 0.118\pm 0.004\;,\;\;\;\;\a0(3\;\mbox{GeV})= 0.42\pm 0.03\;,
\eeq
with $\cd = 4.0/7$. The change in $\a0$ implies that
$\ae(2.5\;\mbox{GeV}) \simeq 0.2\pm 0.1$, which
is reasonable, the perturbative value being around 0.3.
The change in the overall power correction is small (about 5\%),
since, as explained above, the $\mI$-dependence mostly cancels
in Eq.~(\ref{Fpow}).

Using the value (\ref{a0}) of $\a0$, obtained by fitting the thrust
data, one can now predict the power corrections to other event shapes.
For the mean value of the $C$-parameter, for example, we find that
the coefficient $a_F$ in Eq.~(\ref{Fpow}) is
\beq\label{aC}
a_C = 6\,C_F = 8\;.
\eeq
At present there are only data on $\VEV{C}$ at
$Q=M_Z$, where Eqs.~(\ref{a0}) and (\ref{aC}) imply
\beq
\VEV{C}^\pw = 0.057\pm 0.008\;.
\eeq
The second-order perturbative prediction is [\ref{lepqcd}]
\beq
\VEV{C}^\pt = 1.375\,\as + 3.88\,\as^2 = 0.214\pm 0.011\;
\eeq
for $\as = 0.117\pm 0.005$. Thus the full theoretical prediction
is
\beq
\VEV{C}^\th = \VEV{C}^\pt+\VEV{C}^\pw  = 0.271\pm 0.014\;,
\eeq
which is consistent with the experimental result [\ref{ALEPHshapes}]
\beq
\VEV{C}^\ex = 0.2587\pm 0.0013\pm 0.0018\;.
\eeq
Note that the power correction represents over 20\% of the
value of this quantity.

Finally, for the longitudinal cross section fraction $\sL/\st$
we predict a coefficient
\beq
a_L = C_F = 1.33\;,
\eeq
leading to the power correction at $Q=M_Z$
\beq\label{AL}
(\sL/\st)^\pw = 0.010\pm 0.001\;.
\eeq
The first-order perturbative prediction is $\as/\pi = 0.037$.
However, the second-order correction is not yet known.
The preliminary OPAL measurement [\ref{OPALfrag}] is
\beq
(\sL/\st)^\ex = 0.067\pm 0.008\;.
\eeq
Clearly the second-order perturbative correction should be
significant if there is to be satisfactory agreement between
theory and experiment.

The values for $\a0$ obtained above from event shapes are in reasonable
agreement with those deduced from heavy quark fragmentation spectra. In
Ref.~[\ref{DKTheavy}] the value $\a0(2\;\mbox{GeV})\simeq 0.6$
was obtained from fits to heavy quark energy losses in $\ee$
annihilation.  The same conclusion follows from an analysis of the
quantity $\xi_H = -\ln\VEV{x_H}$, where $x_H$ is the energy fraction
carried by the heavy quark $H$, using the approach of the present paper.
We find a quark mass ($1/M$) correction of the form (\ref{Fpow}),
with $Q$ replaced by $M$, $\mR\sim M$, and coefficient $a_H = C_F/2$.
The perturbative prediction deduced from Ref.~[\ref{DKTheavy}] is
\beeq\label{xipert}
\xi_H^\pt
&=& \frac{4C_F}{3\pi} \Biggl\{ \int_M^Q \frac{dk}{k}\,\as(k)
  -  \frac{35}{24} \as(Q)  +   \frac{13}{24} \as(M)\nonumber \\
& & +\;\frac{1}{\b0}(K+\delta_2)\left[\, \as(M) - \as(Q)\,\right]
\Biggr\}\;,
\eeeq
where $\delta_2$  is the (numerically negligible)
2-loop anomalous dimension correction
\beq
 \delta_2 = \left( \frac{53}{18} - \frac{\pi^2}{3}\right) C_F
+ \frac{31}{36} (C_A-2C_F) \>=\> -0.173\;.
\eeq
The expression (\ref{xipert}), which accounts for the
$\as\ln (Q/M)$,  $\as$ and $\as^2 \ln (Q/M)$ terms in $\xi_H$,
but neglects $\as^2$ terms with no large logarithm,
gives $\xi_b^\pt = 0.26\pm 0.02$ for $b$-quarks at $Q=M_Z$.
Comparing with the experimental value of $0.36\pm 0.02$ deduced
from lepton spectra [\ref{Behnke}], this implies that
$\xi_b^\pw = 0.10\pm 0.03$ and hence that
$\a0(2\;\mbox{GeV})\simeq 0.6\pm 0.1$. The errors are
estimated conservatively, taking into account the small
scale $M_b\sim 5$ GeV and the lack of a complete
${\cal O}(\as^2)$ calculation of $\xi_b^\pt$.

\newpage
\noindent{\large\bf 3. Conclusions}

\par\vskip 1ex\noindent
Note that the power correction coefficients $a_T$,
$a_C$ and $a_L$ deduced above using a $\kt$ cutoff
are identical to those obtained in Ref.~[\ref{hadro}]
with a gluon mass cutoff. With a $\kt$ cutoff, however,
these coefficients have a physical interpretation: they
measure the contribution of the low-scale region in which
$\as$ departs significantly from its perturbative form.
After being used to calculate the coefficients, the
cutoff is replaced by an infrared matching parameter
$\mI$, which represents the scale below which we
switch from the perturbative
to the non-perturbative description of $\as$.
As long as $\mI$ is not too small (larger than
about 1 GeV) the predictions are quite insensitive
to its value, indicating that the perturbative
behaviour has set in at that scale.

The divergence in the perturbative expression for
$\as$ at low scales is responsible for the
divergence of the perturbation series for quantities
like those considered here, giving rise to the
so-called ``renormalon ambiguity".
By assuming an infrared regular form for the
effective coupling, we resolve this ambiguity, at
the price of introducing the non-perturbative parameters
$\ap$. These parameters are, however, universal, and can be
measured experimentally, like $\a0$ in Eq.~(\ref{a0}).

Combined fits to the non-perturbative parameters $\ap$
and the perturbative parameter $\as$, using data on
several different event shapes, provide the possibility
of understanding something new about QCD at low scales
and at the same time measuring $\as$ with improved
precision. This would be useful not only for QCD but
also in constraining physics beyond the Standard Model.

\newpage
\noindent{\large\bf References}

\begin{enumerate}
\item\label{DKTheavy}
       Yu.L.\ Dokshitzer, in {\em Proc.\ XXVII Recontre de Moriond:
       Perturbative QCD and Hadronic Interactions, March, 1992},
       ed.\ J.\ Tran Than Van (Editions Frontieres, 1992);
       Yu.L.\ Dokshitzer, V.A.\ Khoze and S.I.\ Troyan, Lund preprints
       LU TP 92--10, 94--23; Yu.L.\ Dokshitzer, in
       {\em Proc.\ Int.\ School of Subnuclear Physics, Erice, 1993}.
\item\label{hadro}
       B.R.\ Webber, Phys.\ Lett.\ B339 (1994) 148;
       see also {\em Proc.\ Summer School on Hadronic Aspects
       of Collider Physics, Zuoz, Switzerland, August 1994},
       ed.\ M.P.\ Locher (PSI, Villigen, 1994).
\item\label{KS}
       G.P.\ Korchemsky and G.\ Sterman, \np{437}{415}{95}.
\item\label{mueller}
       A.H.\ Mueller, in {\em QCD 20 Years Later}, vol.~1
       (World Scientific, Singapore, 1993).
\item\label{CS}
       H.\ Contopanagos and G.\ Sterman, \np{419}{77}{94}.
\item\label{MW}
       A.V.\ Manohar and M.B.\ Wise, \pl{344}{407}{95}.
\item\label{rendis}
       X.\ Ji, preprint MIT-CTP-2381 (hep-ph/9411312).
\item\label{renhvy}
       I.I.\ Bigi, M.A.\ Shifman, N.G.\ Uraltsev and A.I.\ Vainshtein,
       \pr{50}{2234}{94}; M.\ Beneke and V.M.\ Braun, \np{426}{301}{94}; \\
       M.\ Neubert and C.T.\ Sachrajda, preprint CERN-TH.7312/94
       (hep-ph/9407394).
\item\label{renconf}
       U.\ Aglietti and Z.\ Ligeti, preprint CALT-68-1982 (hep-ph/9503209).
\item\label{thrust}
       E.\ Farhi, \prl{39}{1587}{77}.
\item\label{cpar}
       R.K.\ Ellis, D.A.\ Ross and A.E.\ Terrano, \np{178}{421}{81}.
\item\label{sigl}
       P.\ Nason and B.R.\ Webber, \np{421}{473}{94}.
\item\label{CMW90}
       S. Catani, G. Marchesini and B.R. Webber, \np{349}{635}{91}.
\item\label{askt}
       Yu.L.\ Dokshitzer, D.I.\ Dyakonov and S.I.\ Troyan,
       Phys.\ Reports 58 (1980) 270; \\
       D.\ Amati, A.\ Bassetto, M.\ Ciafaloni, G.\ Marchesini
       and G.\ Veneziano, \np{173}{429}{80}.
\item \label{lepqcd}
       Z.\ Kunszt, P.\ Nason, G.\ Marchesini and B.R.\ Webber, in
       {\em Z Physics at LEP 1}, CERN Yellow Book 89-08.
\item\label{glasgow}
       B.R.\ Webber, in {\em Proc.\ XXVII Int.\ Conf.\ on High
       Energy Physics, Glasgow, 1994},\\
       ed.\ P.J.\ Bussey and
       I.G.\ Knowles (Institute of Physics, 1995).
\item\label{ALEPHshapes}
       ALEPH Collaboration, D.\ Buskulic et al., \zp{55}{209}{92}.
\item\label{OPALfrag}
       OPAL Collaboration, D.R.\ Ward, Nucl.\ Phys.\ B (Proc.\ Suppl.)
       39B, C (1995) 134.
\item\label{Behnke}
       T.\ Behnke, in {\em Proc.\ XXVI Int.\ Conf.\ on High Energy
       Physics, Dallas, 1992}, ed.\ J.R.\ Sanford (AIP, 1993).
\end{enumerate}
\end{document}